\documentclass[12pt]{article}
\usepackage{amsmath,amssymb}

\def\bea{\begin{eqnarray}}
\def\ben{\begin{equation}}
\def \een{\end{equation}}
\def \eea{\end{eqnarray}}

\def \nn {\nonumber} 

\def \p{\partial}

\def\beq{\begin{eqnarray}}
\def\eeq{\end{eqnarray}}

\newcommand{\bg}{\begin{gather}}

\newcommand{\bseq}{\begin{subequations}}
\newcommand{\eseq}{\end{subequations}}

\def\half{\frac{1}{2}}

\begin{document}

\title{\textbf{Lifting the Eisenhart-Duval Lift to a Minimal Brane}}

\vspace{2cm}
\author{\bf G.W. Gibbons}
\date{ }
\maketitle
\begin{center}
\hspace{-0mm}
{\bf  D.A.M.T.P.}
\hspace{-0mm}
\end{center}
\begin{center}
\hspace{-0mm}
 {\bf University of Cambridge}
\hspace{-0mm}
\end{center}
\begin{center}
\hspace{-0mm}
{\bf Wilberforce Road, Cambridge  CB3 0WA }
\hspace{-0mm}%
\end{center}
\begin{center}
\hspace{-0mm}
{\bf U.K }
\hspace{-0mm}
\end{center}
\vspace{2cm}


\begin{abstract}
  The motion of a dynamical system on an $n$-dimensional
  configuration space may be regarded as the lightlike shadow
  of null geodsics moving in an $(n+2)$ dimensional spacetime
  known as its Einsenhart-Duval lift. In this paper  it is shown
  that if the configuration space is $n$-dimensional Euclidean
  space, and in the absence of magnetic type  forces, the
  Eisenhart-Duval  lift may be regarded as  an $(n+1)$-brane
  moving in a flat $(n+4)$ -dimensional 
  space with two times. If the  Eisenhart-Duval  lift
  is Ricci flat, then the $(n+1)$-brane moves in such a way as to
  extremise its spacetime volume. A striking example is
  provided by the motion of $N$ point particles moving
  in three-dimensional Euclidean space 
  under the influence of their mutual gravitational  attraction.  
  Embedding with curved configuration space metrics
  and velocity dependent forces  are  also be constructed.
  Some of the issues arising  from the  two times are addressed.
\end{abstract}

\vskip 2 cm
\noindent
\rule{7.7 cm}{.5 pt}\\
\noindent 
\noindent
\noindent ~~~ {gwg1@damtp.cam.ac.uk}

\newpage
 \tableofcontents
\pagebreak

\newpage

\section{The Eisenhart-Duval Lift}

The Eisenhart-Duval lift \cite{Eisenhart,Duval:1984cj,Duval:1990hj}
 has proved itself a powerful tool
for ``geometrising'',  and thus elucidating,
the symmetries of, Lagrangian    dynamical systems
\cite{Cariglia:2014ysa,Cariglia:2016oft}.
In the simplest case one starts from
a Lagrangian of the form  
\beq
L= \half \dot {\bf x} ^2 - V({\bf x} ,t) \,, \label{system}
\eeq
where ${\bf x}$ is a position vector in $\mathbb{E}^n$
with components $x^i$, 
and regards  solutions of the Euler-Lagrange equations
of (\ref{system}) as corresponding to the null geodesics
of an $n+2$ Lorentzian spacetime with coordinates
$X^\mu = (x^i,t,s)\,,i=1,2,\dots,n$ 
admitting an $n+2$ dimensional  Bargmann structure,
that is   covariantly constant null killing vector field
$\frac{\p}{\p s}$.

The Lorentzian metric takes the form
\beq
g_{\mu \nu} dX^\mu dX^\nu = 2Ldt^2 + 2 dt ds =
d {\bf x}^2 + 2 dt ds - 2V({\bf x},t) dt ^2
\label{Elift} 
\eeq
which may be recognized as a Brinkmann
or pp-wave \cite{Brinkmann:1925fr,Ehlers:1962zz}. 
In particular, if for every  $t$,   $V({\bf x},t)$  is a harmonic function,
then the metric $g_{\mu \nu}$ is Ricci flat. That is,
a plane fronted gravitational wave with parallel rays  \cite{Ehlers:1962zz}.
If $V({\bf x},t)$ is both harmonic and quadratic in ${\bf x}$
we have plane gravitational waves with  partially  broken Carroll
symmetry \cite{Duval:2017els} which may be used to model
the gravitational waves currently being detected by LIGO
\cite{Zhang:2017rno,Zhang:2017geq}. The coordinates $t$ and $s$ may be thought
of as Newtonian and Carrollian time respectively \cite{Duval:2014uoa}.

In addition to elucidating the symmetries, of and
equivalences between, dynamical systems
the Eisenhart-Duval construction offers the prospect of applying
global techniques
in  Lorentzian geometry and causal theory  to the  dynamical systems
which are their  shadows  \cite{Minguzzi:2006gq,Minguzzi:2012cx}.

\section{Isometric Embedding of the  Eisenhart-Duval Lift
into a flat space with two times}

An immediate  generalisation of a result valid for $n=3$ 
\cite{Collinson} is that the $n+2$  dimensional Lorentzian
spacetime with metric (\ref{Elift})    
 may be isometrically embedded into flat space
$\mathbb{E} ^{n+2,2}$  equipped with the flat metric  
\beq
 \eta_{AB}dZ^A dZ^B =
d Z^i dZ ^i   - (dZ^{n+1})^2  +(dZ^{n+2}) ^2 + (dZ^{n+3})^2
  -(dZ^{n+4} )^2 \label{flat} 
\eeq  
 $A=1,2,\dots ,n+4$, whose signature is $n+2,2$  
as a co-dimension 2  Lorentzian submanifold, i.e. as an
$9n+1)$-brane \footnote{The
 word ``p-brane'' in its most general sense refers
to an object of $p$-dimensional spatial extent.} 
with world volume coordinates $( {\bf x},s,t )$,  
whose  induced metric coincides with
(\ref{Elift}).  We need  not  assume any special property of
$V({\bf x},t)$.

The embedding is given explicitly by 
$Z^A=Z^A ( {\bf x}, t,s) $, $A=1,2,\dots , n,n+1,n+2,n+3,n+3 $ where 
\bea
Z^i &=& x^i\,,\qquad i=1,2,\dots, n  \nn\\ 
Z^{n+1}&=& \frac{1}{\sqrt{2}} (tV({\bf x},t) +s + t)\,, \nn\\
Z^{n+2}&=& \frac{1}{\sqrt{2}} (tV({\bf x},t)  +s - t )\,,\nn \\
Z^{n+3}&=& \frac{1}{\sqrt{2}}  (V({\bf x},t) + \half t^2) \,,\nn\\ 
Z^{n+4}&=& \frac{1}{\sqrt{2}} (V({\bf x},t) -  \half t^2) \,. \label{embed}
\eea
On may check that if one substitutes (\ref{embed}) into the flat metric
(\ref{flat}) one obtains the Eisenhart-Duval (\ref{Elift})  lift of the   
dynamical system (\ref{system}).
\section{Equation of motion}

We now enquire whether our  $(n+1)$-brane satisfies the
Euler-Lagrange equations arising from extremizing the world-volume.
It is traditional in much of the geometry literature to refer such surfaces
as ``mininal'' without prejudice as to whether the extremum does actually
minimize  the world-volume, hence its use in my title. It certainly
seems to me better than ``on-shell'' for example.  
Bearing that proviso in mind, we vary the action functional
\beq
S[Z^A(x^\mu)]= \int \sqrt{-g} \,d^{n+2} \, x 
\eeq
with respect to the $n+4$ world-volume scalar fields $Z^A(x)$. 
\bea
\delta S&=& \half \int \sqrt{-g} g^{\mu \nu}
\delta (\eta _{AB} \partial_\mu Z^A \partial _\nu Z^B )\, d^{n+2}\,  x \\
&=&  \int \sqrt{-g} g^{\mu \nu} \eta _{AB}
\partial_\mu Z^A \partial _\nu \delta Z^B \, d^{n+2}\, x \\ 
&=-& \int \partial_\nu ( \sqrt{-g} g^{\mu \nu} \eta _{AB}
\partial_\mu Z^A)  \delta Z^B \, d^{n+2}\, x \qquad
        +{\quad boundary \quad term} 
\eea
Thus the Euler-Lagrange equations are
\beq
\partial_\nu ( \sqrt{-g} g^{\mu \nu} \eta _{AB}
\partial_\mu Z^A) =0 \,,
\eeq
which is equivalent that the embedding coordinates $Z^A$ are harmonic
functions with respect to the world-volume metric.
In out case $\sqrt{-g}=-1 $,  therefore we  require  
\bea
0&=&  \partial _\mu \bigl(  g^{\mu \nu}  \partial_\nu Z^A \bigr )  \\ 
&=& \Bigl \{  \frac{\partial^2}{\partial {\bf x} \cdot  {\bf x} } -
2\frac{\partial ^2 }{\partial t \partial s }
+ 2V({\bf x},t) \frac{\partial ^2}{\partial s ^2} \Bigr \}  Z^A =0 \,. 
\eea

Thus in order to satisfy the equations of motion of an $(n+1)$-brane
it suffices that $V({\bf x},t)$ is harmonic in the $n$ vector ${\bf x}$.
That is, if (\ref{Elift}) is  a Ricci flat pp-wave.

An especially appealing example is the case of 
the Newtonian equations of motion of  $N$ point particles in $\mathbb{E}^3$
for which $n=3N$. 

\section{Newtonian Cosmology}

Although often thought of as a quintessentially
general relativistic effect, the Slipher-Lemaitre-Hubble
expansion may be described (at least for distance small
compared with the Hubble radius, i.e. for or small redshifts)
using Newtonian point particle mechanics (see  \cite{Gibbons:2013msa}
and references therein). 
For $N$ point self-gravitating particles with position vectors
${\bf r}_a$, $a=1,2,\dots, N$,
and positive masses $m_a$, a suitable Lagrangian is
\beq
L_{\rm Newton} = \half \sum_{1\le a \le N}  m_a {\dot {\bf r}_a}  ^2 +
\sum_{1\le  a< b \le N} \frac{Gm_am_b }{|{\bf r}_b - {\bf r}_a   | }  \,. \label{Lag} \eeq
To convert (\ref{Lag}) to the form (\ref{system}) we set
\beq
 {\bf x}= ( \sqrt{m_1}{\bf r}_1,  \sqrt{m_2}{\bf r}_2, \dots,
        \sqrt{m_N}{\bf r}_N )  
\eeq
Then
\beq
\frac{\partial^2 V({\bf x},t)  }{\partial {\bf x}  \cdot  \partial {\bf x}}
= \sum _{ 1\le a \le N  } 
-\frac{1}{m_a}  \frac{\partial^2}{\partial {\bf r}_a  \cdot
  \partial {\bf r}_a }   \Bigl( \sum_{1\le  a< b \le N} \frac{Gm_am_b }{|{\bf r}_b - {\bf r}_a   | }      \Bigr ) =0 \,.    
\eeq

One may also treat perturbations around a background
flat Friedmann-Lemaitre model with scale factor $S(t)$
using the Dmitriev-Zeldovich equations \cite{Ellis:2014sla}   
For which a suitable Lagrangian is
\beq
L_1= \half \sum_{1\le a \le N}  m_a  \big( {\dot {\bf r}_a}  ^2  + \frac{\ddot S}{S} {\bf r}_a^2 \bigr )    +
\sum_{1\le  a< b \le N} \frac{Gm_am_b }{|{\bf r}_b - {\bf r}_a   | }  \,. \label{Lag} \eeq
In this case the Eisenhart-Duval lift is no longer  Ricci flat. 

In fact one may define a comoving coordinates ${\bf y}_a$ by
\beq
{\bf r}_a= S(t) {\bf y}_a \,. 
\eeq
The equations in terms of the comoving coordinates  ${\bf y}_a$
may be obtained from a Lagrangian which differs from (\ref{Lag})
by a total time derivative 
\beq
L_2= \half S(t)^2 \sum_{1\le a \le N}  m_a {\dot {\bf y}_a}  ^2 +
\frac{1}{S(t)} \sum_{1\le  a< b \le N} \frac{Gm_am_b }{|{\bf y}_b - {\bf y}_a   | }
\,. \label{DZ}
\eeq

One might wonder whether $L_1$ or $L_2$ give rise to
an Eisenhart-Duval lift which lifts to a brane in higher dimensions.
We leave that question for future study. Another intriguing possible
connection is with \cite{Cariglia:2015dha}. 

\section{Curved Configuration spaces}

It is possible to replace Euclidean $\mathbb{E}^n$  space by 
a curved configuration space $\{Q,g_{ij}\}$ with time independent
metric $g_{ij}(x^k)$ and the discussion
goes through almost unchanged. The embedding
equations (\ref{embed}) provide an embedding into the metric product
$Q \times \mathbb{E}^{2,2}$. Of course it may be that
the configuration space admits an isometric embedding
into some higher dimensional flat space $\mathbb{E}^{u,v}$
in which case we have a brane in $\mathbb{E}^{u+2,v+2}$.

A further extension is obtained by the addition of a velocity dependent
term to the Lagrangian so that
\beq
L= \half g_{ij}(x^k)  \dot x^i \dot x^j + A_i(x^k) \dot x^i -V(x^k,t)  
\eeq
and the lift becomes 
\beq
g_{\mu \nu} dX^\mu dX^\nu = - \bigl(2 V + g^{rs}A_rA_s  \bigr )
  dt ^2 + 2 dt ds + g_{ij} (dx^i + g^{ir} A_r dt)(dx^j + g^{js} A_s dt )\,.  
\eeq
If we define
\beq
\tilde V= V + \half + g^{rs}A_rA_s
\eeq
we have define   
\beq
\tilde Z^{n+1}&=& \frac{1}{\sqrt{2}} (t \tilde V({\bf x},t) +s + t)\,, \nn\\
\tilde Z^{n+2}&=& \frac{1}{\sqrt{2}} (t \tilde V({\bf x},t)  +s - t )\,,\nn \\
\tilde Z^{n+3}&=& \frac{1}{\sqrt{2}}  (\tilde V({\bf x},t) + \half t^2) \,,\nn\\ 
\tilde Z^{n+4}&=& \frac{1}{\sqrt{2}} (\tilde V({\bf x},t) -  \half t^2) \,. \label{embed}
\eeq
and lift the Eisenhart-Duval lift  to $Q \times \mathbb{E}^{2,2}$
with a twisted metric product 
\ben
g_{ij} (dx^i + g^{ir} A_r dt)(dx^j + g^{is} A_s dt ) -
 (dZ^{n+1})^2  +(dZ^{n+2}) ^2 + (dZ^{n+3})^2
 -(dZ^{n+4} )^2 \,.
\een

\section{Discussion}

Using the fact that  one may regard the motion of any
natural mechanical system defined on an $n$-dimensional,
configuration space as   the shadows of    null geodesics moving
in an $(n+2)$ dimensional  Lorentzian spacetime, we have shown that
in the case that the the configurations space
is $n$-dimensional Euclidean space, the Eisenhart-Duval lifted spacetime
may be isometrically embedded in  an $(n +4)$ dimensional
flat space  with two time coordinates.  
The  Eisenhart-Duval  lift is then the history of an $(n+1)$-
brane moving in an ambient flat space
with two time coordinates. In the particular case case that
the Eisenhart-Duval  lift is
Ricci flat, the $n+1$ is a ``minimal''  surface: it  extremises
the spacetime volume of the $(n+1)$ brane. In other words it satisfies
the equations of motion arising from the generalisation to
$(n+1)$ dimensions of the Nambu-Goto action
of string theory. 

That the embedding requires two time coordinates
may seem worrying from a physical point of view
but, as pointed out by Penrose \cite{Penrose:1965rx}  many years  ago, is mathematically inevitable
because the Eisenhart lift is a Bargmann manifold
which  may be regarded  as  gravitational wave.
If they   satisfy   the standard energy conditions, such spacetimes cannot be globally-hyperbolic
and hence cannot admit a time function.
However if the embedding were into a  higher dimensional
Minkowski spacetime with just one time, the intersections with parallel  spacelike hyperplanes would   
provide a time function.

Embeddings into flat spaces with two times   are in fact not unusual. The Ur-example is of course
$(n+1)$-dimensional anti-de-Sitter spacetime. This isometrically embeds
as  a quadric in
$\mathbb{E}^{n,2}$. It has closed timelike curves
(CTC's) and  is not
globally-hyperbolic.  Even if one eliminates the CTC's by passing  to its
universal covering spacetime,
it remains non-globally-hyperbolic

In fact it is possible to isometrically embed more
complicated supergravity solutions representing self-gravitating branes rather  than the test or probe branes
considered in this paper,
into  flat spacetimes with two times. In particular BPS branes.
In this case the embedding  covers the horizon and hence embeds a  non-globally-hyperbolic spacetime
\cite{Gibbons:1998th,Andrianopoli:1999kx, Andrianopoli:1999fn}.
As pointed out in \cite{Tipler:1986qu} the intuitively appealing picture of
strings, or
branes splitting and joining, changing  their
topology as they do so is incompatible with an everywhere non-singular
Lorentzian world  sheet or world volume metric . At the level of classical
solutions, examples show that the resolution of this paradox is that the
world sheet or world volume can remain a smooth submanifold of
Minkowski spacetime, but allow the induced metric to change signature
\cite{Gibbons:2004dz}. That does not happen in the examples exhibited above,
but changes of signature of various types can certainly be expected for
branes moving in a spacetime with more than one time.

At the quantum level, S-matrix amplitudes are typically calculated in
practice by considering Riemann surfaces embedded in Euclidean
space as a function of external Euclidean momenta  and then analytically
continued to timelike   Lorentzian momentum. This is fine
for S-matrix elements but one then faces the problem
of what to do in the context of  cosmology.

For many years Itzhak Bars has been advocating
an approach to the symmetries
of string theory and supergravity based on two  time physics
\cite{Bars:2000qm,Bars:2010zz,Araya:2013bca} .
It would be surprising if there were not some relation of these considerations
to the present paper.  Eisenhart-Duval lifts with  two times
have already been shown 
to arise for higher derivative systems \cite{Galajinsky:2016zer}. 
If they can be isometrically embedded in higher dimensions even more
time dimensions are likely to be required.

Finally let me remark that very recently  the authors of
\cite{Bergshoeff:2020xhv} found, using branes,  a  relation  between
the Carroll and Galilei groups which is differs  from that
in \cite{Duval:2014uoa}. An interesting question is how this fits in
with the construction of the present paper.

\section{Acknowledgements}

The author thanks Marco Cariglia, Peter Horvathy and Paul Townsend
for helpful comments on the manuscript.

\end{document}